\documentclass[preprint2,twoside]{hwo}

\usepackage{lipsum}
\usepackage{natbib}

\newcommand{\kms}{{\rm \, km\, s^{-1}}}

\input{hwo.h}

\begin{document}

\title{\textbf{\LARGE Supermagnified Stars in Lensing Clusters and Small-Scale
 Structure in the Dark Matter}}
\author {\textbf{\large Gabriel Torralba,$^1$ Jordi Miralda-Escud\'e,$^{1,2}$ }}
\affil{$^1$\small\it Institute of Cosmos Sciences, Universitat de Barcelona,
 Barcelona, Catalonia, Spain}
\affil{$^2$\small\it Instituci\'o Catalana de Recerca i Estudis Avan\c cats,
 Barcelona, Catalonia, Spain}



\begin{abstract}
 Supermagnified stars, discovered by HST and JWST in lensing clusters
of galaxies, are luminous stars in source galaxies lying near lensing caustics
that are magnified by large factors ($\sim 1000$), making them detectable at
cosmological distances. Intracluster stars in the lens modify the large-scale
caustic into a corrugated network of micro-caustics, causing frequent
microlensing events of the supermagnified stars. The frequency and lightcurves
of these micro-caustic crossings are extremely sensitive to any small-scale,
low amplitude surface density fluctations in the lensing cluster, making them
a unique probe to dark matter minihalos or any small-scale irregularities. 
Furthermore, the disks of supermagnified stars can be resolved
(at cosmological distances!) from photometric monitoring of micro-caustic
crossing events, exploiting the limb darkening effect. Examples are shown of
model lightcurves to fit the Kelly et al.\ (2018) observations of the first
supermagnified star discovered. The unique sensitivity and angular resolution
of the Habitable Worlds Observatory enables photometry of microlensing events
of supermagnified stars sensitive to small-scale structure different from
intracluster stars, a probe to the nature of dark matter that is not
accessible to other observations at present.
  \\
  \\
\end{abstract}

\vspace{2cm}

\section{Introduction}

 Gravitational lensing offers two unique opportunities to observational
cosmology: exploring the mass distribution in the massive objects
deflecting the light, and analyzing the lensed background sources with
the increased sensitivity and angular resolution provided by the lensing
magnification. Lensing is our only guaranteed direct way to observe the
dark matter (in addition to other direct measurements of acceleration on
objects such as stars), because General Relativity implies that all
forms of matter and energy must interact gravitationally.

 The maximum lensing magnification occur when a source crosses a caustic
curve \citep[e.g.,][]{schneider92,congdon18}. Briefly, caustics are the
lens mapping of critical curves from the image plane back to the source
plane; critical curves are the curves in the image plane where the
determinant of the magnification matrix,
$A=\partial {\bf y} / \partial {\bf x}$, is equal to zero.
We refer to the image
plane as the plane in the sky we observe after the lensing deflection,
and the source plane as what we would observe in the absence of lensing.
An angular position $\bf x$ in the image plane is mapped into a position
${\bf y}({\bf x})$ in the source plane by tracing the gravitational
deflection of a light ray backwards. 
At the critical curves, the mapping of the image plane onto the source
plane reaches an extremum, which means that when the source crosses a
caustic from outside to inside, two new images appear on each side of
the critical line.

 Generally, the separation $\delta x$ of these two new images from the
critical line varies with the separation $\delta y$ of the source from
the caustic as $\delta x \sim (\delta y)^2$ (except at special points
called cusps that deviate from the generic fold caustic).
Consequently, the magnification varies as
$\mu = | {\rm det}\, A |^{-1} \propto (\delta x)^{-1}\propto
(\delta y)^{-1/2}$. The magnification of a source
moving at a constant angular velocity behind the lens, from the outside
to the inside of a caustic, is predicted to fall as
$\mu \propto (\delta t)^{-1/2}$, after a time
$\delta t$ since the caustic crossing; when moving from inside to
outside, the magnification will instead increase as
$(-\delta t)^{-1/2}$ until the two images disappear at the caustic
crossing. The maximum
magnification $\mu_{\rm max}$ reached at the caustic crossing is
determined by three limits: (1) real light sources are always extended
with some angular size $\theta_s$,
so $\mu_{\rm max}\propto \theta_s^{-1/2}$; (2) diffraction reduces the
maximum magnification derived from geometric optics when the wavelength
is $\lambda > D_l (\theta_s \mu_{\rm max})^2$; (3) any small-scale
structure in the mass distribution breaks up the lens caustic of a
smoothed model of the mass distribution into a corrugated network of
microcaustics, with a reduced maximum magnification.

 The most massive lenses in the Universe are clusters of galaxies, with
typical deflection angles and critical line sizes of
$\theta_l \sim 20''$ or $10^{-4}$ radians. If their mass distribution
were smooth the maximum magnification of a source star with
$\theta_s\sim 10^{-16}$ (the angular size of a star of $10 R_\odot$ at
redshift $z\sim 1$) could be as high as
$\mu_{max}\sim (\theta_l/\theta_s)^{1/2} 10^6$,
allowing for many stars in a distant galaxy to be detected by HST
\citep{miralda91}. However, galaxy clusters are full of
intracluster stars that were tidally stripped from their parent galaxies.
Near the {\it macro-caustic} of a smooth model of the cluster lens, a
very small fraction of the mass in the form of point masses such as
intracluster stars is enough to turn the macro-caustic into a
corrugated band of much smaller micro-caustics.

 As we discuss below, this break-up of the cluster macro-caustic into
many micro-caustics has the disadvantage of reducing the maximum
magnification, thereby limiting the detection of individual stars to
only the most luminous stars in a galaxy. However, the advantage is
that microcaustic crossings cause short-time variability of the
magnification that facilitates the identification of what we designate
as ``supermagnified stars''.

 The first supermagnified star was discovered by \citet{Kelly18} in the
cluster MACS J1149; it is a blue supergiant and one of the most luminous
stars in a source galaxy at $z_s=1.49$. Since then, more than a dozen
of cases have been identified
\citep[e.g.,][]{CKD19,WCD22,DPK22,CKT22,MZJ23,DMA23,MCZ23,DSY23};
one of the most spectacular cases is the
large number of candidates for supermagnified stars in the ``Dragon
arc'' produced by a galaxy at $z_s=0.725$ in the cluster Abell 370
\citep{Fudamoto25}.

\section{Micro-caustics from Intracluster Stars}
\label{sec_micro}

 The properties of the corrugated band of microcaustics produced by
the population of intracluster stars in a lensing cluster were discussed
in detail in \citet{VDM17} (see also the recent review of
\cite{Weisenbach24}). A summary of the typical characteristics is
as follows:
\begin{enumerate}
 \item The cluster critical line becomes a band of micro-critical lines
of width $\theta_w \sim \kappa_\star/g$, where $\kappa_\star$ is the
surface density of microlenses divided by the critical surface density,
and $g$ is the gradient of the magnification matrix eigenvalue that
vanishes at the critical line in the smooth model.
A typical cluster has $\kappa_\star \sim 0.005$ as inferred from the
surface brightness of the intracluster star population,
$g^{-1}\sim 20''$, and $\theta_w\sim 0.1''$.
 \item The spacing between micro-critical lines in the corrugated band
is $\theta_l \sim \theta_\star \kappa_\star^{-1/2}$, where
$\theta_\star$ is the Einstein radius of an isolated intracluster star.
The number of micro-critical lines crossed when moving across the band
is $N_c \sim \theta_w/\theta_l$. For typical microlenses,
$\theta_\star \sim 1\, \mu{\rm as}$, $\theta_l\sim 10\, \mu{\rm as}$,
and $N_c\sim 10^4$.
 \item The corrugated band of micro-caustics has a width in the source
plane $\theta_c \sim \kappa_\star^2/g$, which is typically
$\theta_c \sim 500\, \mu{\rm as}$. A source star moving with a typical
transverse velocity $v_t\sim 500 \kms$ will take $\sim 10^4$ years to
move over the whole band of micro-caustics, with an average crossing
rate of 1 micro-caustic per year. The rate of micro-caustic crossing
is, however, not uniform, but follows the $(\delta t)^{-1/2}$ profile
from the smooth-model caustic crossing.
 \item The typical maximum magnification at each micro-caustic crossing
is $\mu_{\rm max} \sim (\theta_\star/\theta_s)^{1/2} \kappa_\star^{-3/4}$,
or $\mu_{\rm max} \sim 10^3$ for a red supergiant angular size
$\theta_\star\sim 10^{-3}\, \mu{\rm as}$ and other typical values
mentioned previously. At a micro-caustic crossing, the magnification of
the two brightest micro-images is usually higher than the sum of
magnifications of all the other micro-images.
\end{enumerate}

 The numbers given as examples above need to be understood only as
typical quantities; in reality, they of course have a distribution
and can vary substantially depending on various lensing parameters.

 The recent finding of many variable point sources, likely to be
supermagnified stars, in the Dragon arc in A370 \citep{Fudamoto25}
demonstrates the potential of JWST for discovering many more examples
of this phenomenon. Most supermagnified stars are near the limit of
detection and are probably identified only during one of the
magnification peaks during a micro-caustic crossing. A few brighter
stars, however, may be seen continuously as they traverse the band of
micro-caustics and experience multiple peaks at each individual
micro-caustic crossing. This is the case for the first supermagnified
star named Icarus, which experienced a peak magnification in 2016 that
seems consistent with a micro-caustic crossing.

\begin{figure*}[ht]
\begin{center}
\includegraphics[width=0.7\textwidth]{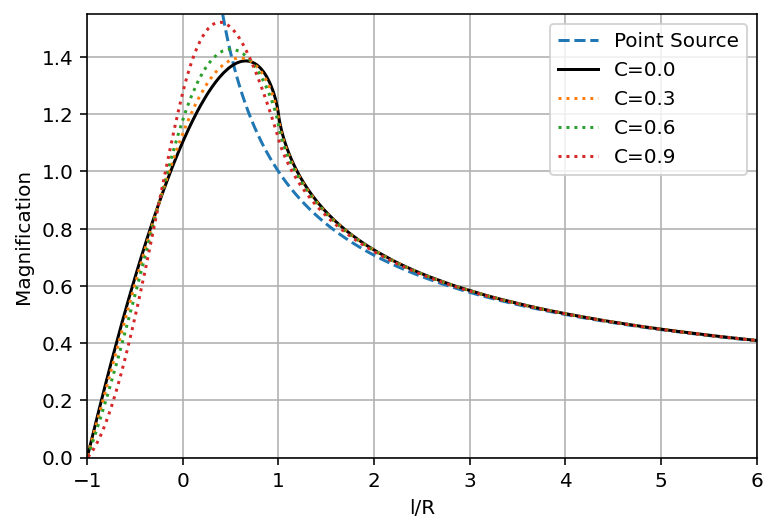}
\caption{\small Magnification $\mu$ of a star crossing a micro-caustic,
as a function of the distance from the star center to the micro-caustic,
$\ell$, divided by the stellar radius $R$. A point source has the
lightcurve $\mu = (\ell/R)^{-1/2}$ (dashed blue line) and disappears at
$\ell < 0$; the uniform disk model is shown as the solid black line,
with a lightcurve that rises linearly from zero at $\ell > -R$.
Including limb darkening modeled as a linear decrease of surface
brightness with projected radius results in the dotted curves, with
increasing fraction of surface brightness decline from the center to the
edge ($C$).
\label{fig1_label}
}
\end{center}
\end{figure*}

\section{Predicted lightcurves at micro-caustic crossings: Impact of
Limb Darkening}

 The promise of monitoring in detail the lightcurve of a supermagnified
star, in particular when a micro-caustic crossing occurs, is that these
photometric variations are extremely sensitive to low-amplitude,
small-scale variations in the dark matter surface density. During a
micro-caustic crossing, observing the photometric lightcurve is
actually resolving the angular disk of a star at cosmological distance,
at an angular scale $\theta_s$ of $\sim $ few nano-arc second for a red
supergiant, or even smaller for a blue supergiant or one of the most
massive O stars in the main-sequence. Variations on the scale
$\mu \theta_s$ with an amplitude as low as $\mu^{-1}$ produce large
perturbations in the expected lightcurves at micro-caustic crossings.
These variations may be produced, for example, by axion dark matter
minihalos, which are otherwise not possible to detect in lensing because
of their highly subcritical surface density \citep{DM20}.

 In principle, the precise shape of an isolated micro-caustic crossing
(one that is dominated by the magnification profile of the two
micro-images merging at the micro-caustic and not significantly
affected by variations from other neighboring micro-caustics) is
accurately predictable. The simple analytic solution of the lightcurve
for a uniform disk \citep{miralda91} is easily extended to any detailed
limb darkening model, with a profile that depends on the photometric
filter. Unfortunately, many of the most luminous stars in a galaxy,
such as supergiants, have limb darkening profiles that are uncertain
owing to their strong winds, with very thick photospheres. Based on
models for smaller giant stars \citep{NL13}, we adopt a simple linear
form for the limb darkening model where the surface brightness varies
with the ratio of the projected radius on the stellar surface to the
star radius, $\lambda$, as
\begin{equation}
 I(\lambda) = I_0 (1-C \lambda) ~, 
\end{equation}
where $I_0$ is the central surface brightness. Note that the angle of
incidence $\beta$ on the photosphere is given by $\cos\beta=\lambda$.

 The predicted lightcurves for this model are shown in
Figure \ref{fig1_label}.
The dashed blue line is the point source curve and the solid black line
is for the uniform disk ($C=0$). When the separation from the
microcaustic is large, all curves are proportional to $\ell^{-1/2}$,
where $\ell$ is the separation from the micro-caustic. We normalize all
the curves to $\mu = (\ell/R)^{-1/2}$ at large $\ell$, where $R$ is the
radius of the star, implying that the total flux of the star is the
same in all cases. In terms of observed time,
$\ell = v_t(t-t_0)/(1+z_s)$, where $t_0$ is the time when the center of
the star crosses the micro-caustic, and $v_t$ is the transverse velocity
of the source relative to the observer-lens line of sight, projected to
the direction perpendicular to the micro-caustic. Typical values
of this velocity, arising mostly from the bulk large-scale structure
velocities of the Milky Way, the lens and the source, are $v_t\sim
500 \kms$, implying a time scale for the stellar disk to cross the
micro-caustic that ranges from several hours for an O star to a week
for a red supergiant. For negative $v_t$, the lightcurve is reversed,
with two images disappearing instead of appearing at the micro-critical
line.

\begin{figure*}[ht]
\begin{center}
\includegraphics[width=0.7\textwidth]{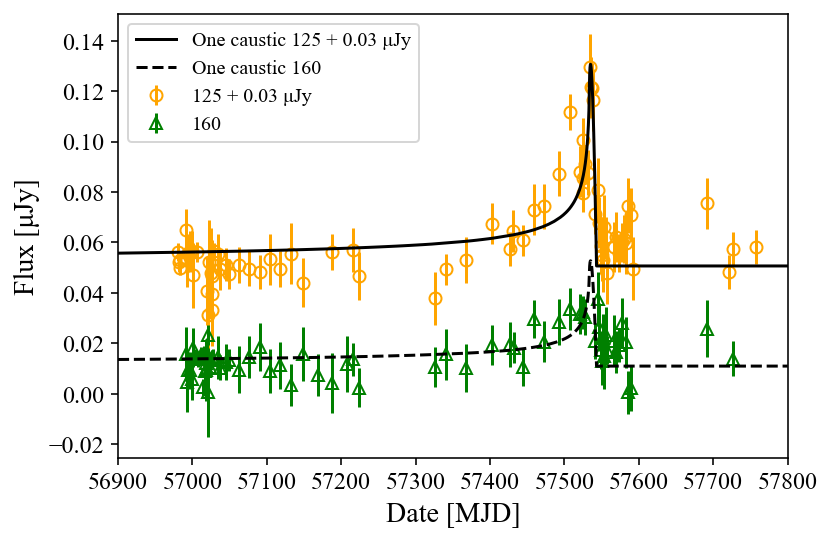}
\caption{\small Observations of the Icarus supermagnified star
\citep{Kelly18} are shown as orange and green triangles with error bars
for the F125W and F160W filters, respectively. The orange points are
raised by $0.03\, \mu{\rm Jy}$ for clarity. A 6-parameter, single
micro-caustic crossing model fit to both filters is shown (solid line
for F125W, dashed line for F160W).
\label{fig2_label}
}
\end{center}
\end{figure*}

\begin{figure*}[ht]
\begin{center}
\includegraphics[width=0.7\textwidth]{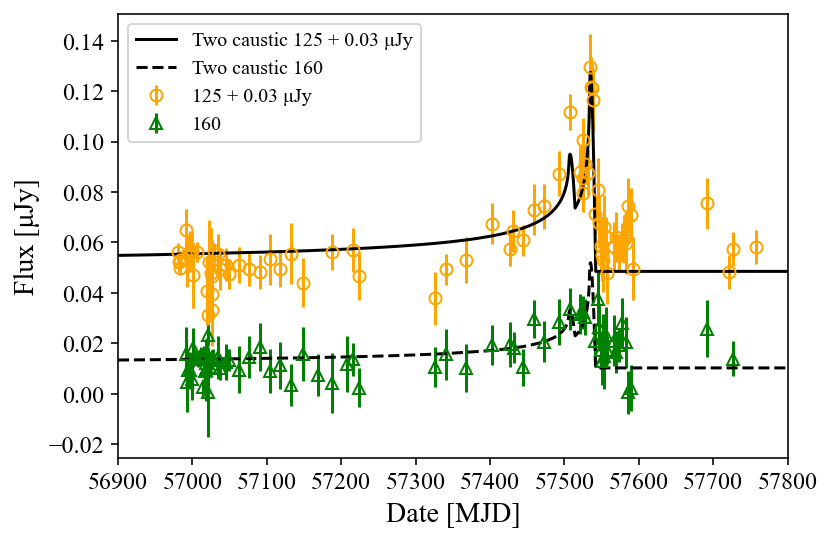}
\caption{\small Same as Figure \ref{fig2_label},
but for a model with two independent
caustics with different positions and amplitudes, with a total of 8 free
parameters for both filters.
\label{fig3_label}
}
\end{center}
\end{figure*}

 The models with limb darkening, ranging from a 30\% ($C=0.3$) to a
90\% ($C=0.9$) drop of the surface brightness from the center to the
limb, show that as limb darkening increases, the lightcurve adopts a
more symmetric shape, with the magnification peak shifting to a more
central position around the times when magnification reaches the
extrapolated point source curve at the stellar radius position
$\ell=R$. The magnification starts rising slower as the stellar edge
enters the micro-caustic, for larger limb-darkening, and then
accelerates the rise. Without limb darkening, the rise when the edge
enters the micro-caustic is more linear, and the peak is asymmetric
with a very steep fall near $\ell/R=1$ when the star moves fully
inside the caustic.

 Although the changes are small, the possibility of accurately
following the photometric lightcurve is exciting because the detection
of limb darkening would demonstrate that the angular disk of a star at
a cosmological distance has not only been resolved, but its profile has
been measured with enough detail that the physics of the photosphere
can be tested. In principle, the limb darkening effect depends on the
photometric filter, and may be inverted (with the limb brighter than the 
center) in some photospheric emission lines. This would provide enough
confidence in the measurements, for the detection of perturbations
from these curves to be strong evidence for the presence of small-scale
irregularities in the dark matter surface density.

\section{Example: fits to the observed Icarus magnification peak}

 The brightest supermagnified star detected so far is still the first
one reported by \citet{Kelly18}, a blue supergiant at $z_s=1.49$ named
Icarus. Although it is usually
at an AB magnitude near 27 at $1.25\, \mu{\rm m}$, it rose to
AB magnitude near 26 in May 2016, or a rise of nearly
$0.1\, \mu{\rm Jy}$. As we have argued in Section \ref{sec_micro},
for typical parameters we expect to see about 1 micro-caustic crossing
per year with a duration of a few days. Hence, the observed event is
likely, but certainly not guaranteed, to be a rather isolated peak from
a single micro-caustic crossing. We therefore try to model it this way,
fitting the expected lightcurve shape described above to the HST
observations of Icarus in two filters, F125W and F160W, the two filters
that have most of the observations.

 Looking at the data shown in Figure 3 of \citet{Kelly18} for the F125W
filter, the first thing we notice is that the duration of the peak
appears to be around 50 days. For a typical transverse velocity
$v_t\sim 500 \kms$ and radius of one of the most luminous blue
supergiants in a star-forming galaxy, $R\sim 1\, {\rm AU}$, the duration
should be only a few days. However, the micro-caustic crossing time
depends only on one component of the transverse velocity, projected to
the plane of the sky and the direction perpendicular to the
micro-caustic, so the probability distribution of $v_t$ is flat at
small $v_t$. We therefore assume that $v_t$ in this lens may be as low
as $\sim 50 \kms$, with a probability of occurring randomly that is
still $\sim 10\%$.

  Assuming a single-caustic crossing, and fixing initially $C=0$, we fit
the data of the two filters with four parameters in each filter: the
micro-caustic crossing time, the peak duration (equal to
$R(1+z_s)/v_t$), the maximum flux at the magnification peak, and a
baseline flux that remains constant and is added to the flux of the
micro-images merging at the micro-critical line. Note that this constant
baseline flux may be either from the rest of the micro-images of Icarus,
or from any associated unresolved light (such a star cluster that Icarus
might be part of). When doing the joint fit to the data in the two
filters, we assume that the crossing time and the peak duration are the
same but the peak and baseline flux are different, so we have a total
of six parameters for fitting the two filters. We use the UltraNest
\footnote{\url{https://johannesbuchner.github.io/UltraNest/}}
package of \citet{Buchner21}, based on the nested sampling Monte Carlo
algorithm MLFriends \citep{Buchner16}.

 Figure \ref{fig2_label} shows the fit to a single caustic crossing.
Orange and green circles with error bars are the \citet{Kelly18}
observations used for the fit, in the F125W and F160W filters,
respectively. The orange points for F125W are raised by
$0.03\, \mu{\rm Jy}$ to help better visualize the data for the two
filters together. The total number of points is 131, and with 6 free
parameters we expect $\chi^2=125 \pm 16$ for a good fit. The best
fit that is found is shown as the solid curve for the orange points,
and the dashed curve for the green points, and has a
value of $\chi^2=184.4$. This value indicates the fit is not perfect
but comes close to reproducing the observations within the reported
error bars.

 Allowing the limb darkening parameter $C$ to vary, we find there are
no substantial variations of $\chi^2$ with $C$, meaning that the
photometric precision is in this case too low to detect any limb
darkening effect, with a variation of $\chi^2$ of less than unity. 
Nevertheless, improved photometry of these events with JWST and,
hopefully, HWO in the future, may provide cases of greatly improved
photometric precision where these effects may become detectable.

  Although there may be systematic uncertainties in the photometry that
may account for the higher $\chi^2$ than expected, we try also the
possibility of fitting to two independent micro-caustic crossings that
happen to be close together. In this case we add a new crossing time
and peak magnification for the second caustic crossing, but keep the
duration and baseline flux to be the same for the two caustics, so the
number of free parameters is increased to 6 per filter. As before, the
joint fit for the two filters assumes the same crossing times for the
two caustics and peak duration (3 parameters), but different baseline
fluxes (2 parameters) and peak fluxes; however, the ratio of the peak
fluxes of the two micro-caustic crossings is the same for each filter
(because we assume the same star crosses different micro-caustics at
the same transverse velocity), so the peak fluxes add another
3 parameters. In total, we have 8 free parameters in this case.

  The result of the 2 micro-caustic fit is shown in
Figure \ref{fig3_label}. The total $\chi^2$, now expected to be
$123\pm 16$, is reduced to $165.5$. Although the reduction is
sufficient to make this a better model, the fit is not yet very good
and the second caustic improves the fit mostly by reducing the
discrepancy with the F125W high point near MJD 57500. Again, no
significant dependence on the $C$ parameters for the two filters is
found. These results will be presented in more detail in an upcoming
paper (Torralba \& Miralda-Escud\'e, in preparation).

 We note that our assumption that isolated crossings of fold
micro-caustics are more likely than more complex configurations may not
necessarily be correct: for example, for a broad distribution
of masses of the intracluster stars, micro-caustics arising from red
and brown dwarfs could be crossed more frequently.

\section{Conclusions and Prospects for HWO Science}

  With our present capabilities for performing photometry on the
faintest sources we can observe with HST and JWST, we are discovering
the cases of supermagnified stars arising from the most luminous stars
in a typical star-forming galaxy, when they reach magnifications in the
thousands. These stars offer a unique probe to small-scale structure
in the dark matter that is otherwise difficult to observe. Microlensing
of the few supermagnified stars discovered so far can already set
interesting limits to any putative population of compact objects that
may be part of the dark matter \citep{Oguri18,Muller25}, such as
primordial black holes. But beyond that, the lightcurves of
micro-caustic crossings are uniquely sensitive to low-amplitude
fluctuations in the dark matter surface density at these small scales.
The example of minihalos predicted in theories of axion dark matter
was explored in \citet{DM20}, but there are other cases where dark
matter structures that may reveal to us the true nature of dark matter,
and are usually too diffuse to come close to the critical surface
density necessary to induce multiple imaging and large magnifications
on background sources, can be detectable in the lightcurves of
supermagnified stars.

 The superb photometric sensitivity and angular resolution that is
planned for the {\it Habitable Worlds Observatory} should make it the
most powerful instrument to observe these micro-caustic crossings and
explore in this way the nature of dark matter. If the HWO cameras can
make full use of the telescope diffraction limit in the blue band of
$\sim 10\, {\rm mas}$, this will cleanly separate the flux of the
supermagnified star from other background light in its host galaxy and
the lensing cluster, yielding photometry limited only by the number of
detected photons from the point source of the magnified star. While
JWST, with its focus in the infrared, works best for following red
supergiants, HWO would be our most powerful instrument for following
supermagnified blue supergiants and O main-sequence stars at relatively
low redshift, where the brightest cases of supermagnification are
likely to be discovered. O stars can also be followed in the
ultraviolet where most of their luminosity is emitted, and their small
sizes may present important advantages for detecting microlensing
structure on small scales.

 Supermagnified stars in the visual filters may also be observed with
larger telescopes from the ground, such as the Extremely Large
Telescope, which can reduce photometric noise by colleting more photons
with their aperture larger than HWO. The photometric precision that can
be reached on the faint supermagnified stars with these telescopes
depends on how well adaptive optics can perform, and the systematic
uncertainties in subtracting the complex host galaxy optical structure.



\acknowledgements
{\bf Acknowledgements.} We are grateful to Liang Dai for scientific
discussions. The authors acknowledge support by the Spanish AEI
grant PID2022-137268NB-C52 funded by MCIN/AEI/10.13039/501100011033/FEDER.

\bibliography{miralda}

\end{document}